\documentclass[twocolumn,superscriptaddress,showpacs,preprintnumbers,amsmath,amssymb,aps,pre]{revtex4}
\usepackage{graphicx}
\usepackage{dcolumn}
\usepackage{bm}
\usepackage{color}
\begin{document}
\title{Geoelectric field and seismicity changes 
preceding the 2018 M$_w$6.8 earthquake and the subsequent activity in Greece}
\author{N. V. Sarlis}
\affiliation{Solid State Section and Solid Earth Physics Institute, Physics Department, University of Athens, Panepistimiopolis, Zografos 157 84,
Athens, Greece}
\author{E. S. Skordas}
\affiliation{Solid State Section and Solid Earth Physics Institute, Physics Department, University of Athens, Panepistimiopolis, Zografos 157 84,
Athens, Greece}
\author{P. A. Varotsos}
\email{pvaro@otenet.gr}
\affiliation{Solid State Section and Solid Earth Physics Institute, Physics Department, University of Athens, Panepistimiopolis, Zografos 157 84,
Athens, Greece}

\begin{abstract}
A strong earthquake of magnitude M$_w$6.8 struck Western Greece 
on 25 October 2018 with epicenter at 37.515$^o$N 20.564$^o$E. It was preceded by an anomalous geolectric signal
 that was recorded on 2 October 2018 at a measuring station 70km
away from the epicenter. Upon analyzing  
this signal  in natural time, we find that it conforms to the 
conditions suggested (e.g., Entropy {\bf 2017}, 19, 177)
for its identification as precursory Seismic Electric 
Signal (SES) activity. Notably, the observed lead time of 23 days lies within the range of values that has been very
recently identified (Entropy {\bf 2018}, 20, 561) as 
being statistically significant for the precursory  
variations of the electric field of the Earth. 
Moreover, the analysis in natural time  of the seismicity subsequent to 
the 
SES activity in the area candidate to suffer this strong earthquake reveals that 
the criticality conditions were obeyed early in the morning of 18 October 2018, i.e., 
almost a week before the strong earthquake occurrence, in 
agreement with earlier findings. {Furthermore,  upon  
employing the recent
method of nowcasting earthquakes, which is based on natural time, 
we find an earthquake potential score around 80\% just before the occurrence 
of this M$_w$6.8 earthquake. Here, we also report the recording 
of more recent SES activities including a very recent one which just appeared at Pirgos 
measuring station on 13 October 2023.}
\end{abstract}

\pacs{05.40.-a, 05.45.Tp, 91.30.Dk, 89.75.-k}
\maketitle
\section{Introduction}\label{intro}
\begin{figure}
\centering
\includegraphics[scale=0.45]{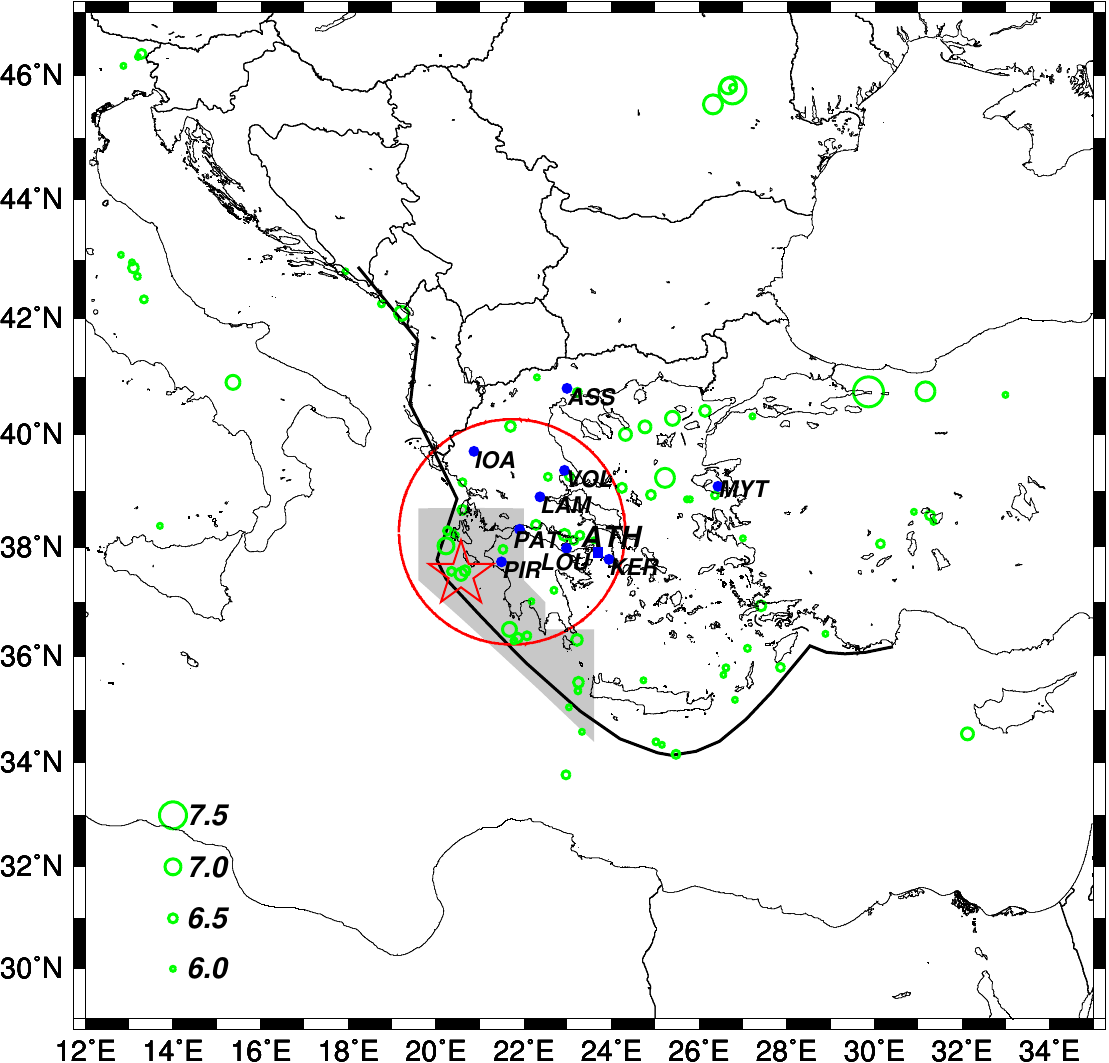}
\caption{(color online) Map of the larger area $N_{29}^{47}E_{12}^{35}$ in which 
the EQs of magnitude greater than or equal to 6.0 are shown by the 
green circles. The locations of the measuring stations operating in 
Greece of the VAN telemetric network are shown by the blue 
circles. The thick black line depicts the Hellenic arc\cite{DOL99} while the gray 
shaded area the selectivity map of Pirgos (PIR) measuring station. 
The red star corresponds to the epicenter of the M$_w$6.8 EQ on 
25 October 2018 and the red circle delimits a circular region 
with radius $R=225$km around the city of Patras.  \label{fig:1}}
\end{figure}

According to the United States Geological Survey (USGS) \cite{MOU18}, 
a strong earthquake (EQ) of moment magnitude M$_w$6.8 occurred on 25 
October 2018 22:55 UTC at an epicentral distance around 133 km SW of 
the city  of Patras, Western Greece. Patras has a metropolitan area 
inhabited by a quarter of a million persons and fatal casualties 
have been probably avoided because, among others, at 22:23 UTC 
almost half an hour before the strong EQ 
a moderate EQ of magnitude M=5.0 occurred approximately at the 
same area as the strong EQ\cite{LIT18} (see Fig.\ref{fig:1}).
 
\begin{figure}
\centering
\includegraphics[scale=0.33]{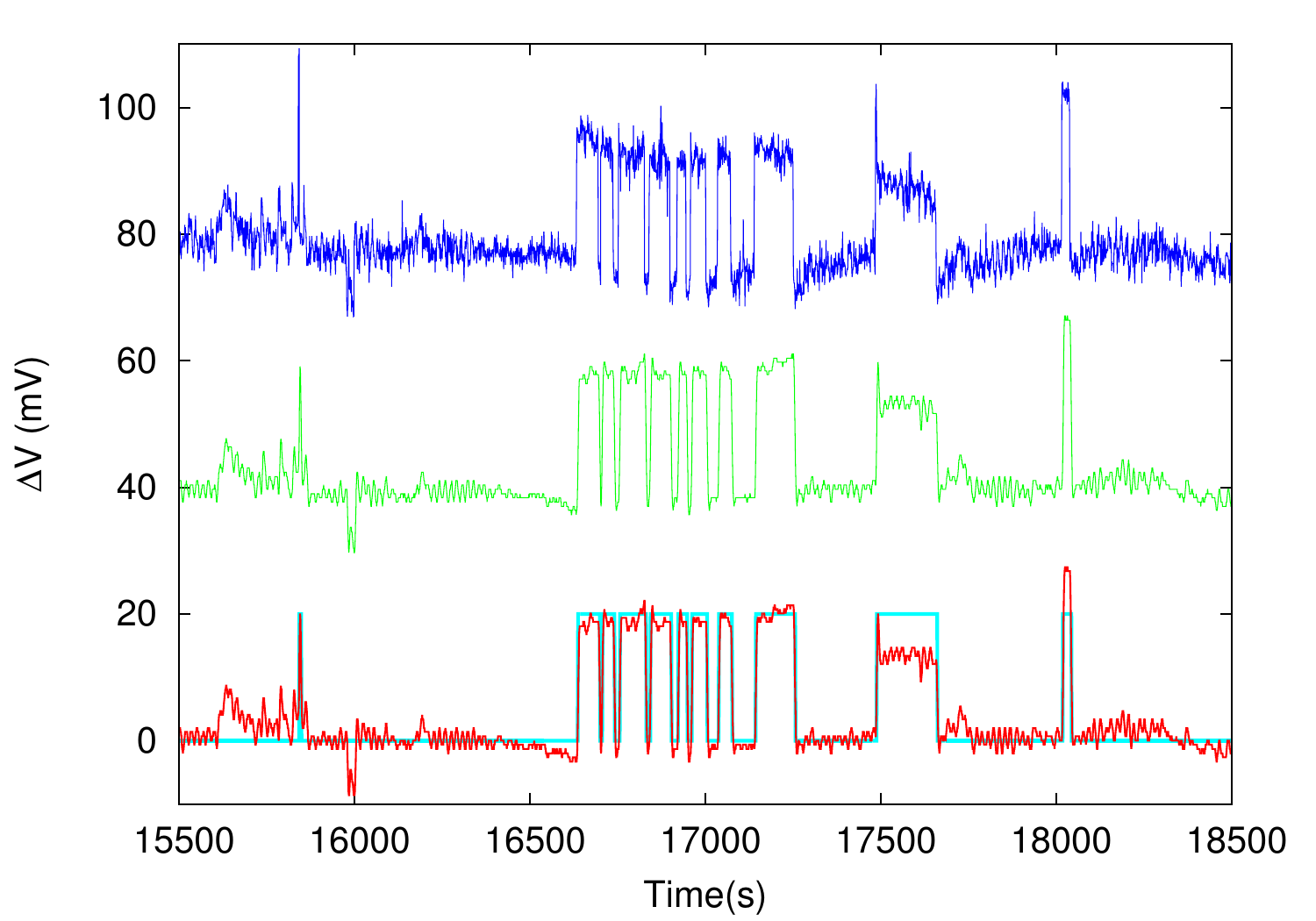}
\caption{(color online) The SES activity recorded at three (almost parallel)
 measuring dipoles at  PIR  station. 
 The time is measured in seconds since 00:00 UTC on 2 October 2018.
 The dichotomous representation of the SES activity is shown in the 
 bottom channel by the cyan line. \label{fig:2} }
\end{figure}

Geoelectric field continuous monitoring is operating in Greece 
by the Solid Earth Physics Institute\cite{VAR96B,NEWBOOK,SPRINGER}
 at 9 measuring field stations (see the blue circles in Fig.\ref{fig:1}) aiming at 
 detecting Seismic Electric Signals (SES). SES are 
 low frequency ($\le 1$Hz) variations 
of the electric field of the Earth that have been found
to precede strong EQs in Greece\cite{VAR84A,VAR84B,VAR91,VAR93,NAT08}, 
Japan\cite{UYE00,UYE02,ORI12}, China\cite{ZLO01,HUA11,GAO10,FAN10},
 Mexico\cite{RAM08,RAM11},  and elsewhere\cite{SES18}. 
 They are emitted due to the cooperative
orientation of the electric dipoles {\cite{VARBOOK,NEWBOOK,VAR74M}} (that any how exist due 
to defects\cite{LAZ85,VAR08438} in the rocks) of the future focal 
area when the gradually increasing stress before the strong EQ reaches 
a critical value\cite{VAR93}.  SES may appear either as single pulses or 
in the form of SES activities, i.e, many pulses within a relatively short time period ({\cite{VAR91},} e.g. see Fig.\ref{fig:2}).
The lead time of single SES is less 
than or equal to 11 days while for SES activities it 
varies from a few weeks up to 5$\frac{1}{2}$ months {\cite{VAR91,SPRINGER}}. 
SES are recorded\cite{VAR91,VAR93} at sensitive points\cite{V36} on the Earth's surface
which have been selected after long experimentation 
in Greece during 1980s and 1990s 
that led\cite{livan,MARY} to the construction of the {so-called} VAN  telemetric network  (from the acronym of the scientists 
Varotsos, Alexopoulos and Nomicos who pioneered this research) 
comprising the 9 measuring field stations depicted in Fig.\ref{fig:1}. 
Each measuring station records SES from specific {EQ} prone areas 
which constitute the {so-called}  selectivity map of the station {\cite{VAR93,UYE96,DOL99}}. 
The gray shaded area of Fig.\ref{fig:1} depicts the selectivity map for 
the Pirgos (PIR) measuring station as it
resulted after the recording  of SES from various 
epicentral areas {\cite{SAR08}}. A basic criterion for distinguishing SES from noise
 is that the recorded signal should\cite{VAR91} exhibit properties 
 compatible 
with the fact that it was emitted far away from the recording station. 
This is usually called\cite{VAR91}   $\Delta V/L$ criterion (where $\Delta V$
stands for the potential difference between two electrodes that constitute a measuring electric dipole and $L$ for 
the distance between them) and has been 
found\cite{VAR98,SAR99,JAP2,JAP3} to be compatible 
with the aforementioned SES generation model if we take into account
 that EQs occur in faults (where resistivity is usually orders 
 of magnitude smaller than that of the surrounding rocks, e.g., 
 see \cite{NEWBOOK} and  references {therein}).
 
The SES research has been greatly advanced after the 
introduction  of the concept of natural 
time in 2001 {\cite{NAT01,NAT02,NAT02A}}. Firstly, the criticality properties  of SES activities (like the existence of long-range correlations and unique entropic properties) has been revealed by natural time analysis and 
hence new possibilities have been provided for the identification of SES
and their distinction from man-made noise {\cite{NAT03A,NAT03B,NAT05C,NAT06A,NAT06B,NAT08,NAT09V}}. 
Secondly, natural time analysis allowed the 
introduction of an order parameter for seismicity the 
study of which 
allows 
the determination of the occurrence time of the strong EQ  within
 a few days up to one week {\cite{NAT05C,NAT07,SAR08,NEWBK,SPRINGER,Varotsos2015,EQS2}}. Thirdly, minima of the fluctuations of 
 the order parameter of seismicity have been identified
 before all shallow {EQs} with ${\rm M} \geq 7.6$ 
 in Japan during the 27 year period from 1 January 1984 to 11 March 2011, 
 the date of the M9.0 Tohoku EQ occurrence {\cite{PNAS13,PNAS15}}. 
 Finally, the interrelation of SES activities and seismicity has been 
 further clarified because when studying the EQ magnitude time series 
 in Japan it was found that the minimum of the 
 fluctuations of the order parameter of seismicity, which  is observed simultaneously with the initiation of an SES activity \cite{TECTO13}, appears
when long range correlations prevail {\cite{JGR14}}. 

The scope of  {the major part of this paper that 
appeared on 16 November 2018 in Ref.\cite{SARSKO18} by the first two authors was twofold:} 
First, to report the geoelectrical field
 changes (SES) observed before the M$_w$6.8 EQ that occurred on 25 
October 2018. Second, to present the natural time analysis of both 
the SES activity and the seismicity preceding this EQ. 
In this version, we present what happened after this M$_w$6.8 EQ including 
the additional SES activities recorded (Section IV) and updating the nowcasting results 
in subsection III.D a summary of our results and the main 
conclusions are presented in the final Section.

\section{Natural time analysis. Background}
 
Natural time analysis, introduced in the beginning of 2000's
\citep{NAT01,NAT02,NAT02A,NAT03B,NAT03A}, uncovers unique dynamic features hidden behind the time series of complex systems.
In a time series comprising $N$
events, the natural time $\chi_k = k/N$ serves as an index
for the occurrence of the $k$-th event. This
index together with the energy $Q_k$ released  during the $k$-th event, 
i.e., the pair $(\chi_k, Q_k)$, is studied in
natural time analysis. Alternatively, one studies the pair
$(\chi_k,p_k)$, where

\begin{equation}
p_k=\frac{Q_k}{\sum_{n=1}^NQ_n} \label{eq1}
\end{equation}
stands for the normalized energy released during the $k$-th
event. As it is obvious from Eq.(\ref{eq1}),  the correct estimation of $p_k$ simply demands that
$Q_k$ should be proportional to the energy emitted during the $k$-th event. Thus, for SES activities $Q_k$ is proportional to the duration of the $k$-th  pulse while for {EQs} it is proportional to the energy emitted\cite{KAN78} during the $k$-th {EQ} of magnitude $M_k$, i.e., $Q_k \propto 10^{1.5M_k}$  (see also \cite{SPRINGER,SAR17}).
The variance of
$\chi$ weighted for $p_k$, labeled by $\kappa_1$,  is
given by \citep{NAT01,NAT03B,NAT03A,NAT05C,SPRINGER}

\begin{equation}\label{kappa1}
\kappa_1=\sum_{k=1}^N p_k (\chi_k)^2- \left(\sum_{k=1}^N p_k
\chi_k \right)^2.
\end{equation}

For the case of seismicity, the quantity $\kappa_1$  
has been proposed to be an order parameter since  
$\kappa_1$  changes abruptly when a mainshock (the
new phase) occurs, and in addition the statistical properties of
its fluctuations are similar to those in other non-equilibrium and
equilibrium critical systems {(\cite{NAT05C}, see also pp. 249-253 of Ref. \cite{SPRINGER})} . 
{It has been also found that $\kappa_1$ is a key parameter 
that enables recognition
of the complex dynamical system under study entering the critical
stage \citep{NAT01,NAT02,NAT02A,SPRINGER}. This occurs when $\kappa_1$ becomes
equal to 0.070 {(\cite{PNAS}, see also p. 343 of Ref.\cite{SPRINGER})}.  In Table 8.1 of Ref.\cite{SPRINGER} one
can find a compilation of 14 cases  including a variety of dynamical
models in which the condition
$\kappa_1$=0.070 
has been ascertained 
(cf. this has been also later confirmed in the 
analyses of very recent experimental 
results in Japan by Hayakawa and coworkers {\cite{HAY15,POT18,POT18419}}).
 Especially for the case of SES activities, 
it has  been found  that when they are 
analyzed in natural time we find $\kappa_1$ values close 
to 0.070 {(\cite{NAT01,NAT02,NAT03A,NAT03B}, e.g. see Table 4.6 on p. 227 of Ref. \cite{SPRINGER})}, i.e., 
\begin{equation}
\kappa_1\approx 0.070.
\label{eqk1} 
\end{equation}
When analyzing in natural 
time the small EQs with magnitudes greater than or equal to a 
threshold magnitude $M_{thres}$ that occur after the initiation of an SES activity within the selectivity map 
of the measuring station that recorded the SES activity, 
the condition $\kappa_1=0.070$ is 
found to hold for a variety of $M_{thres}$ a few days up to one week
 before the strong EQ occurrence {\cite{NAT05C,NAT06A,NAT07,SAR08,UYE09,NEWBK,SPRINGER,PNAS,Varotsos2015,EQS2} }. This is very important from practical point of view
 because it enables the estimation of the occurrence time of a strong EQ 
 with an accuracy of one week, or so.

The entropy $S$ in natural time is defined\cite{NAT01,NAT03B,NAT04,SPRINGER,SAR17} 
 by the relation
\begin{equation}\label{S}
S=\sum_{k=1}^N p_k \chi_k \ln  \chi_k  -  \left( \sum_{k=1}^N p_k \chi_k \right)  \ln \left( \sum_{m=1}^N p_m \chi_m \right).
\end{equation}
It is a dynamic entropy showing \cite{NAT05B} positivity, 
concavity and Lesche~\cite{LES82,LES04} experimental stability. 
When $Q_k$ are independent and identically-distributed random variables, 
$S$ approaches\cite{NAT04} the value 
$S_u\equiv\frac{\ln2}{2} -\frac{1}{4} \approx 0.0966$ 
that corresponds to the case $Q_k=1/N$, which within the context of natural 
time is  usually termed ``uniform'' distribution {\cite{NAT03B,SPRINGER,SAR17}}.
 Notably, $S$ changes its value to $S_{-}$ upon time-reversal, i.e., when the first event becomes last ($Q_1 \rightarrow Q_N$), the second last but one ($Q_2 \rightarrow Q_{N-1}$) etc,
\begin{eqnarray}
S_{-}& = &\sum_{k=1}^N p_{N-k+1} \chi_k \ln  \chi_k  \nonumber \\
& - &  \left( \sum_{k=1}^N p_{N-k+1} \chi_k \right)  
\ln \left( \sum_{m=1}^N p_{N-m+1} \chi_m \right), \label{Splin} 
\end{eqnarray}
and hence it gives us the possibility to observe the 
(true) time-arrow\cite{NAT05B}. Interestingly, it has been 
established \cite{SPRINGER,SAR17} that both $S$ and $S_{-}$ for SES activities are smaller than $S_u$,
\begin{equation}
S,S_{-}\le S_u. \label{SSU}
\end{equation}
On the other hand,  these conditions are violated
for a variety of similar looking electrical noises {(e.g. see Table 4.6 on p. 228 of Ref. \cite{SPRINGER})}.

Natural time has been recently employed 
by Turcotte and coworkers\citep{RUN16,RUN18A,LUG18A,LUG18B} 
as a basis for a new method to estimate the current level of seismic risk called ``earthquake nowcasting''.
This will be explained in the next Section.

\section{Results}
\subsection{Geoelectric field changes}
Figure \ref{fig:2} depicts an SES activity that
 was recorded in PIR station (see Fig.\ref{fig:1}), which comprises a multitude of measuring 
 dipoles, on 2 October 2018 between 04:20 and 05:05 UTC. The potential differences $\Delta V$ of three of these electric dipoles of comparable length $L$ (a few km) deployed in the NEE direction are shown. The true headings of these dipoles 
 are from top to bottom in Fig.\ref{fig:2} are $75.48^o$, $64.83^o$, and $76.16^o$. An inspection of this figure 
reveals that the SES activity resembles a telegraph 
signal with periods of activity and periods 
of inactivity as it is usually the case {\cite{NAT02,NAT03A,NAT03B}}.  If we impose 
a threshold in the $\Delta V$ variation\cite{NAT02,NAT03A,NAT03B}, we can obtain the dichotomous (0-1) representation of the SES activity  depicted by the cyan color in Fig.\ref{fig:2}.

\subsection{Natural time analysis of geoelectrical signals. Criteria for distinguishing SES}
Apart from the aforementioned $\Delta V/L$ criterion suggested
long ago for the distinction of SES from man-made noise {\cite{VAR91}}, 
natural time analysis has provided, as mentioned, three additional criteria
 for the classification of an electric signal as SES activity. These criteria are Eq.(\ref{eqk1}) and the 
 conditions (\ref{SSU}). The analysis in natural time of the 
 dichotomous representation 
shown in Fig.\ref{fig:2} 
results in $\kappa_1=0.072(2)$, $S=0.066(2)$ and 
$S_{-}=0.079(3)$, which are obviously compatible with the criteria 
for distinguishing SES from noise. This  leads us to support that 
the anomalous variation of the electric field of the Earth observed on 2 October 2018
is indeed an SES activity.

\subsection{Estimation of the occurrence time of the impending EQ}
 We now follow the method suggested in Ref.\cite{SAR08} for the 
estimation of the occurrence time of the impending strong EQ by analyzing in natural time all the small 
EQs of magnitude greater than or equal to $M_{thres}$ 
that occurred after the initiation of the SES activity recorded on 2 October 2018 within 
the selectivity map of PIR measuring 
station shown by the gray shaded area in Fig.\ref{fig:1}. 
The { EQ} catalog\cite{NOA18} of the 
Institute of Geodynamics of the National Observatory of Athens 
has been used and each time a new small 
EQ takes place we calculate the $\kappa_1$ values corresponding to the events that occurred within all the possible subareas of the 
PIR selectivity map that include this EQ {\cite{SAR08}}. This procedure leads to an ensemble of $\kappa_1$ values from 
which we can calculate the probability Prob($\kappa_1$) of $\kappa_1$ to lie within $\kappa_1\pm 0.025$.
Figures \ref{fig:3}(a), (b), (c), and (d) depict the histograms of Prob($\kappa_1$) obtained after 
the occurrence of each small EQ with magnitude ML(ATH) \cite{VAR91,VAR93} greater than or
 equal to M$_{{\rm thres}}$=2.7, 2.8, 2.9, and 3.0, 
respectively. We observe that within a period of 5 hours around 
18 October 2018 00:30 UTC all the four distributions Prob($\kappa_1$) 
exhibit a maximum at $\kappa_1=0.070$. This behavior has been found, as already mentioned, to occur a few 
days up to one week or so before the strong EQ occurrence {\cite{SAR08,NEWBK,SPRINGER,Varotsos2015,EQS2}}. Actually, one week 
later, i.e., on 25 October 2018, a strong M$_w$6.8 EQ occured\cite{MOU18} within the selectivity 
map of the PIR measuring station (see the red star in Fig.\ref{fig:1}). 
Interestingly, as it is written in the legends of the panels of Fig.\ref{fig:3}, 
two of the three small EQs that led to the fulfillment of the 
criticality condition $\kappa_1=0.070$
originated from epicentral areas located only 20 or 25km south from that of the strong EQ.

\begin{figure}
\centering
\includegraphics[scale=0.4]{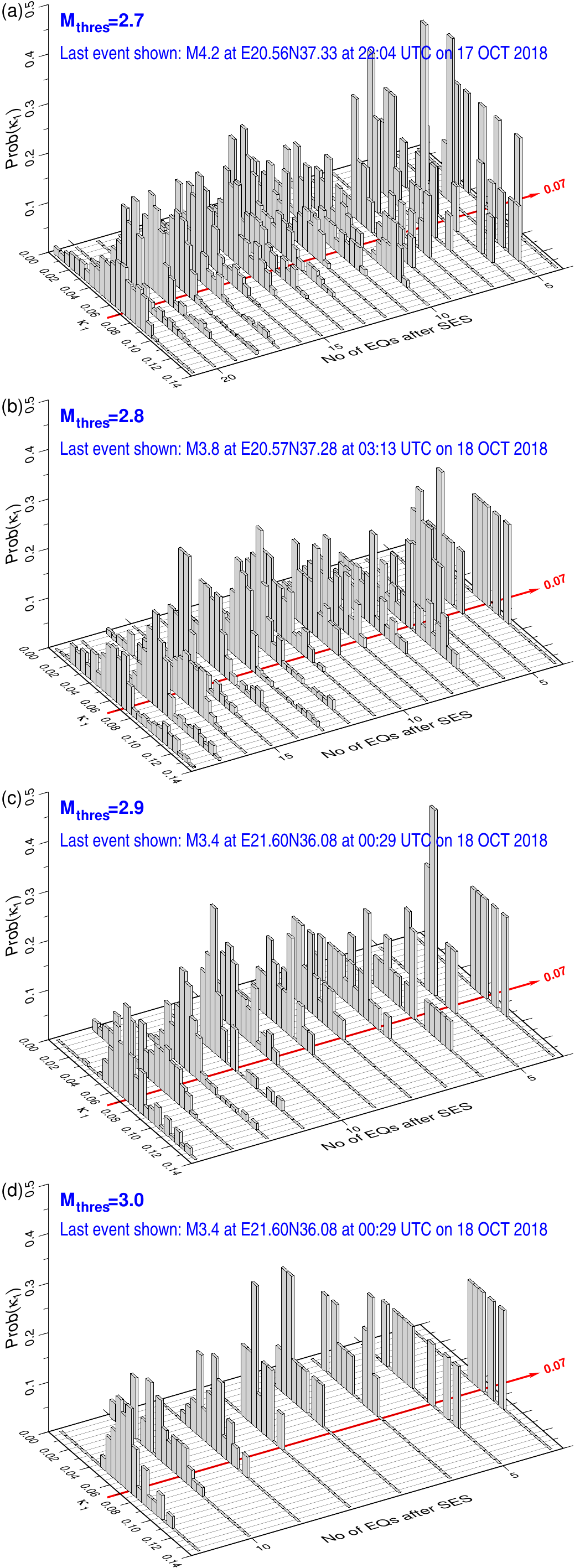}
\caption{(color online) (a) to (d): The probability distribution Prob$(\kappa_1)$ of $\kappa_1$ versus $\kappa_1$  as it results after the occurrence of each small EQ within the selectivity area of PIR (see the gray shaded area in  Fig. \ref{fig:1}) for 
various magnitude thresholds $M_{thres}$=2.7, 2.8, 2.9, and 3.0. \label{fig:3} }
\end{figure}

\subsection{Estimation of the current level of risk 
by applying {EQ} nowcasting}
Nowcasting EQs is a recent method  for the determination of the 
current state of a fault system and the estimation of the current progress in the EQ cycle\cite{RUN16}.
It uses a global EQ catalog to calculate from ``small'' EQs the level of hazard for ``large'' 
EQs. This is achieved by employing the natural time concept and count 
the number $n_s$ of ``small'' EQs that occur after a ``strong'' EQ. 
The current value $n(t)$ of $n_s$ since the occurrence of the last ``strong'' EQ is 
compared with the cumulative distribution function (cdf) $P(n_s<n(t))$ of $n_s$ obtained when ensuring 
that we have enough data to span at least 20 or more ``large'' EQ cycles.  The {EQ}  potential 
score (EPS) which equals the ``current'' cdf value, EPS=$P(n_s<n(t))$  is therefore a unique 
measure of the 
current level of hazard and assigns a number between 0\% and 100\% to every region so defined. 
Nowcasting EQs has already found many useful applications\cite{RUN18A,LUG18A,LUG18B} among which is 
the estimation of seismic risk to Global Megacities. For this application\cite{RUN18A} 
the EQs with depths smaller than a certain value $D$ within a larger area are studied in order  
to obtain the cdf $P(n_s<n(t))$. Then the number $\tilde{n}_s$ of ``small'' EQs 
 around a Megacity, e.g., EQs in a circular region of epicentral distances smaller
 than a radius $R$ with hypocenters shallower than $D$,
 is counted since the occurrence of the last ``strong'' EQ in this  region. Based on the  
 ergodicity of EQs that has been proven\cite{FER99,TIA03,TIA07} by using the metric published 
 in Refs.\cite{MOU89,MOU92}, \citet{RUN18A} suggested that the seismic risk around a Megacity 
 can be estimated by using the EPS corresponding to the current $\tilde{n}_s$ 
 estimated in the circular 
 region. Especially in their Fig.2, they used the large 
 area $N_{29}^{47}E_{12}^{35}$ in order to 
estimate the EPS for EQs of magnitude greater than or equal to 6.5 at an area of radius $R=400km$
 around the capital of Greece Athens. Figure \ref{fig:4} shows the results of a similar 
 calculation based on the  United States National {EQ} Information Center PDE catalog
 (the data of which are available from Ref.\cite{NEICPDE}) which we performed 
 focusing on the city of Patras, Greece, for EQs of magnitude greater than {or} equal to 6.0. 
 Notably before the occurrence of the M$_w$6.8 EQ on 25 October 2018  EPS was 
 found to be as high as 80\%{, see the blue lines in Fig. \ref{fig:4}.
 A similar calculation since the occurrence of the latter M$_w$6.8 EQ
 until 19 January 2019 leads to $\tilde{n}_s=205$ resulting in EPS 78\% 
 (see the green line in Fig. \ref{fig:4}).}

\begin{figure}
\centering
\includegraphics[scale=0.35]{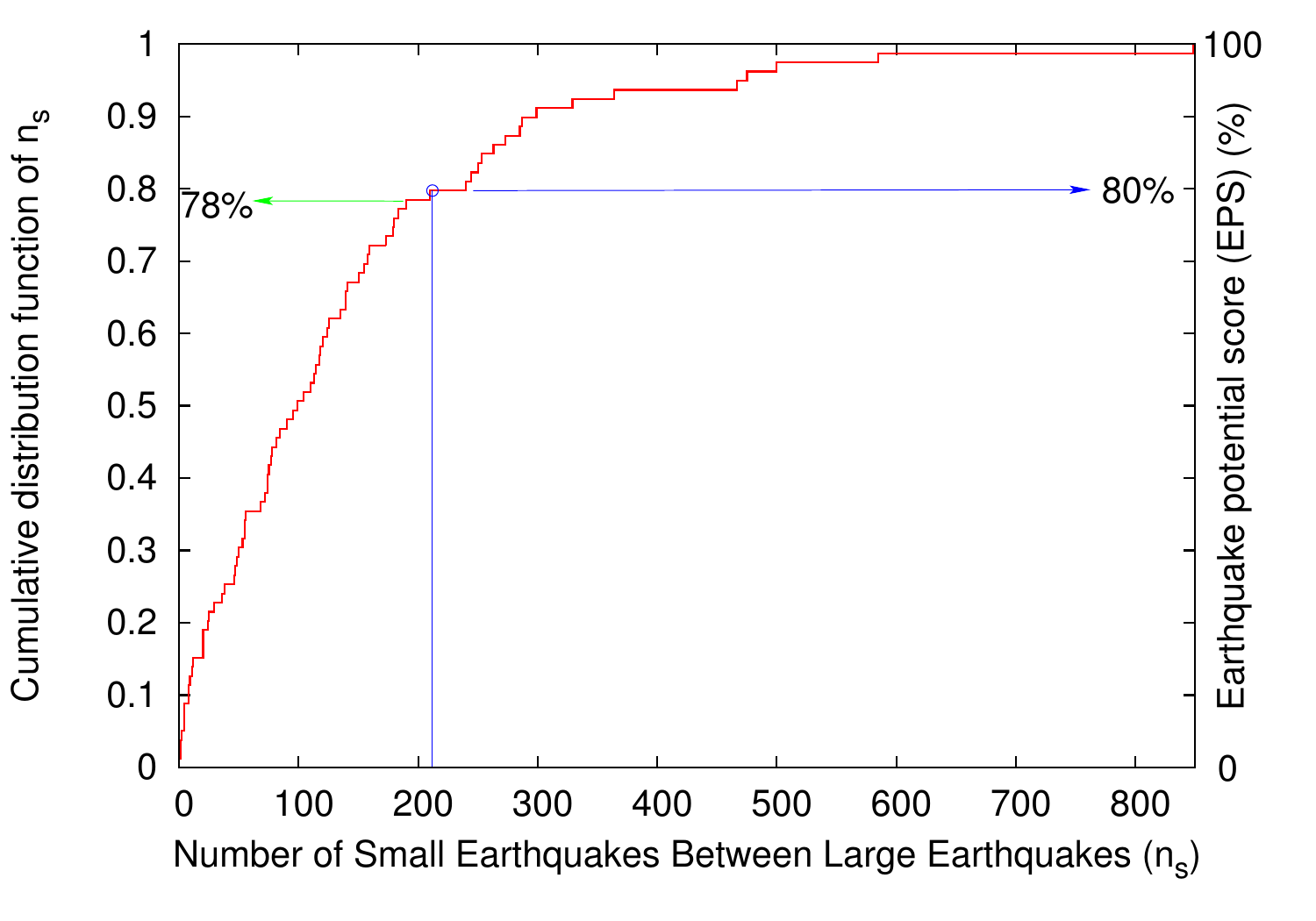}
\caption{(color online)Estimation of the {EQ} potential score (EPS) based 
on the cumulative distribution function of the number
 $n_s$ of the small EQs ($6.0> M\ge 4.0$) that occur within 
 the large area $N_{29}^{47}E_{12}^{35}$  
 depicted in Fig.\ref{fig:1} between the
 occurrence of two strong ($M\ge 6.0$) EQs. The number $\tilde{n}_s$ 
 of small EQs that occurred within $R\le 225$km with 
 depths $D\le 200$km from the city of Patras: 
 (i){since the M$_w$6.5 EQ on 17 
 November 2015 07:11 UTC \cite{NID15} and before 
 the occurrence of the M$_w$6.8 EQ on 25 October 2018 was 212 (see the blue lines for EPS)  
 and  (ii)since the occurrence of the latter EQ on 25 October 2018 until 19 January 2019 is 205 (see the 
 green line for EPS).}
   \label{fig:4} }
\end{figure}

\section{Discussion}

Recently, the statistical significance of the Earth's electric and magnetic field 
variations preceding {EQs} has been studied\cite{SAR18} on the basis of the 
modern tools of event coincidence analysis\cite{DON16,SCH16,SIE17} and receiver 
operating characteristics\cite{FAW06,SARCHRIS14}. Using an SES 
dataset\cite{VAR91,VAR93,DOL93} from 1980s it was found that SES are 
statistically significant precursors to EQs for lead times in the following four distinct time periods:
 3 to 9 days,
 18 to 24 days, 43 to 47 days, and 58 to 62 days {(the first one corresponds to single SES, while the latter three to SES activities \cite{SAR18})}. Since the SES activity, shown in Fig.\ref{fig:2},
 was recorded on 2 October 2018, the SES lead time for the present case of the M$_w$6.8 EQ
 on 25 October 2018, which is 23 days, favorably falls within the second time period of 18 to 24 days. Moreover,
 the analysis of the seismicity subsequent to the initiation of the SES activity in the 
selectivity area of the PIR station has led to the conclusion that the criticality condition $\kappa_1=0.070$
has been satisfied early in the morning on 18 October 2018. This compares favorably with the time window 
of a few days up to one week already found from 
various SES activities in Greece, Japan and United States {\cite{SAR08,NEWBK,UYE09,SPRINGER,Varotsos2015,EQS2}}.

 Let us now turn to the results concerning the entropy of the SES activity of Fig.\ref{fig:2} in natural
 time. As it was reported both $S$ and $S_{-}$ are well below $S_u$ in accordance with the 
 findings (e.g. see Ref. \cite{SAR17}) so far for SES activities. Based on the critical properties 
 that characterize the emission of signals that precede rupture (i.e., infinite range
correlations compatible with a detrended fluctuation analysis (DFA)\cite{PEN94,PEN95,KAN01} 
exponent $\alpha_{DFA}=1$)  a fractional Brownian 
 motion \cite{MAN68,MAN69} model has been 
 suggested\cite{NAT06A} according to which both $S$ and $S_{-}$ values should scatter
around 0.079 with a standard deviation of 0.011 (see Fig.4 of Ref.\cite{NAT06A}). 
Interestingly, the values $S=0.066(2)$ and $S_{-}=0.079(3)$ of the SES activity recorded on 2 October 2018 are fully compatible with this model.

 Finally, the successful results (i.e., the 80\% EPS found before the occurrence of the M$_w$6.8 EQ
 on 25 October 2018) from the {EQ} nowcasting method which is based on natural 
time are very promising. Nowcasting does not involve any model and there are no free parameters to be fit 
to the data {\cite{RUN16}}. 

{On 3 January 2019, an electrical activity (Fig.\ref{fig:5}) was recorded at PAT measuring 
station (close to the city of Patras see Fig.\ref{fig:1}) which was classified as an SES activity because its
natural time analysis by using the procedure described in Ref.\cite{NAT09V}
 resulted in $\kappa_1=0.075(22)$, $S=0.071(22)$, and $S_-=0.075(30)$. Similar 
 conclusion was drawn for an electrical activity recorded at PIR on 9 January 2019. 
 To estimate the occurrence time of the impending EQ, we currently analyze in 
 natural time (as in subsection III.C)  the subsequent seismic activity  occurring 
within the area comprising the gray shaded area of Fig.\ref{fig:1} 
and the one around PAT (in view of 
the green line in Fig.\ref{fig:4}, see also the rectangle with solid lines in 
Fig.8 of Ref.\cite{SAR08}) \cite{On17APR,On19JUN}.  }

\begin{figure}
\centering
\includegraphics[scale=0.38,angle=270]{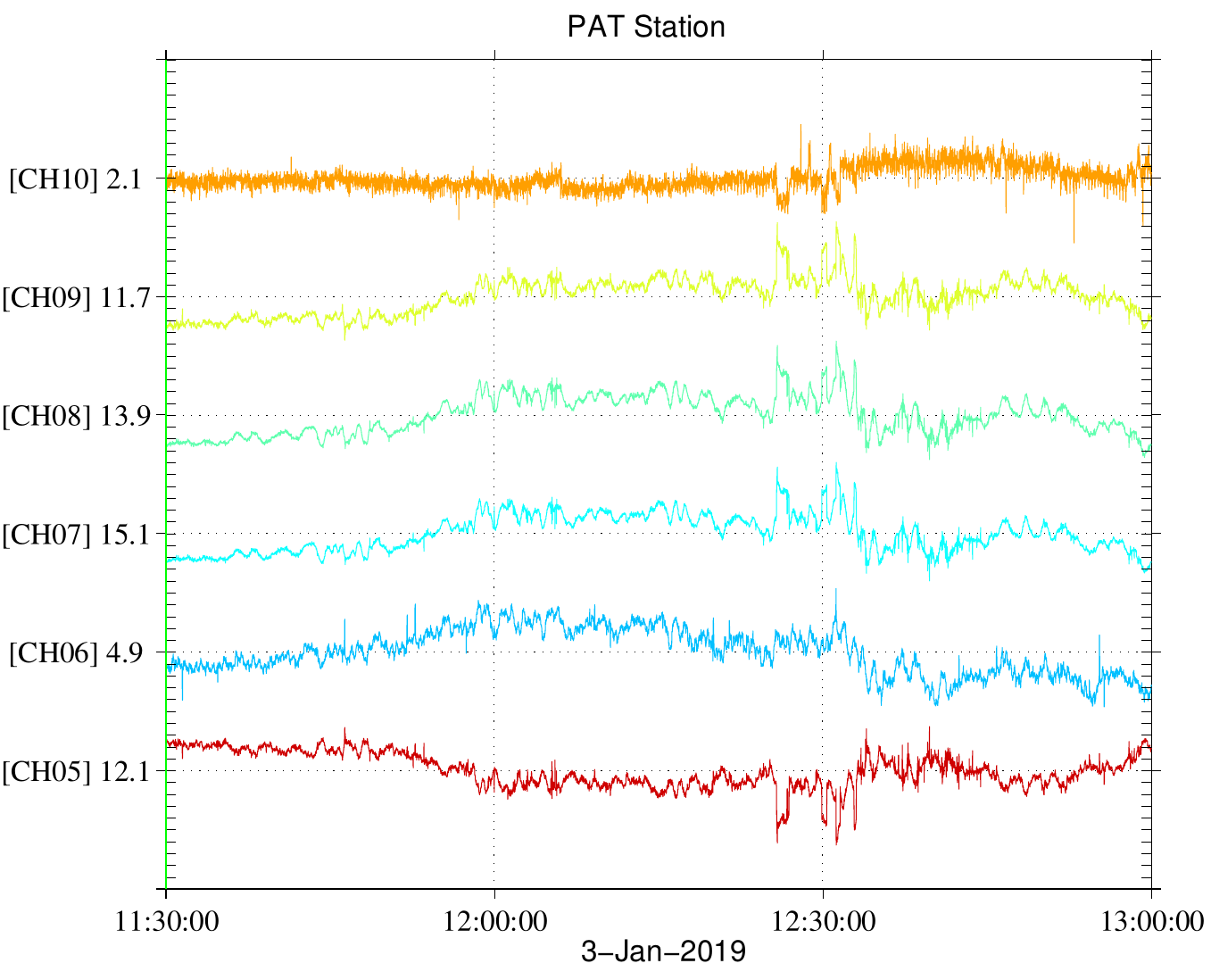}
\caption{{(color online) The SES activity on 3 January 2019 at PAT measuring 
station recorded on a multitude of electric dipoles labeled CH05 to CH10. 
The numbers on the left vertical axis denote the scale  of the variation of the potential 
difference in mV per division for each dipole.}   \label{fig:5} }
\end{figure}

\section{Summary and Conclusions}
The strong EQ of magnitude M$_w$6.8 that occurred in Western Greece 
 on 25 October 2018 was preceded by an SES activity on 2 October 2018 recorded at PIR measuring 
 station of VAN telemetric network. The EQ epicenter was located within the selectivity map of PIR
 depicted by the gray shaded area in Fig.\ref{fig:1}.

 The lead time of 23 days between the precursory SES activity and the strong EQ is 
 statistically significant as recently found by the recent methods of event coincidence 
 analysis and receiver operating characteristics. Both the entropy $S$ and the entropy $S_{-}$ under 
time reversal in natural time are compatible with previous observation for SES activities as 
well as agree with a model for SES activities based on fractional Brownian motion. The analysis in natural time of 
the seismicity subsequent to the SES activity by 
considering the events occurring within the selectivity area of PIR shows that criticality has been 
reached early morning on 18 October 2018, almost 
a week before the strong EQ occurrence in accordance with the earlier findings.
{When 
employing the recent method of nowcasting earthquakes, which is based on natural time, 
we find an earthquake potential score around 80\% just before the occurrence 
of the M$_w$6.8 earthquake on 25 October 2018. Here,  
we also report the recording of more recent
 SES activities at PIR and PAT \cite{On17APR,On19JUN,On24JAN,On17FEB}.}
 
 \acknowledgments
 We gratefully acknowledge the continuous supervision
and technical support of the geoelectrical stations of the
VAN telemetric network by Vasilis Dimitropoulos, Spyros
Tzigkos and George Lampithianakis.

 \begin{figure}
\centering
\includegraphics[scale=0.6]{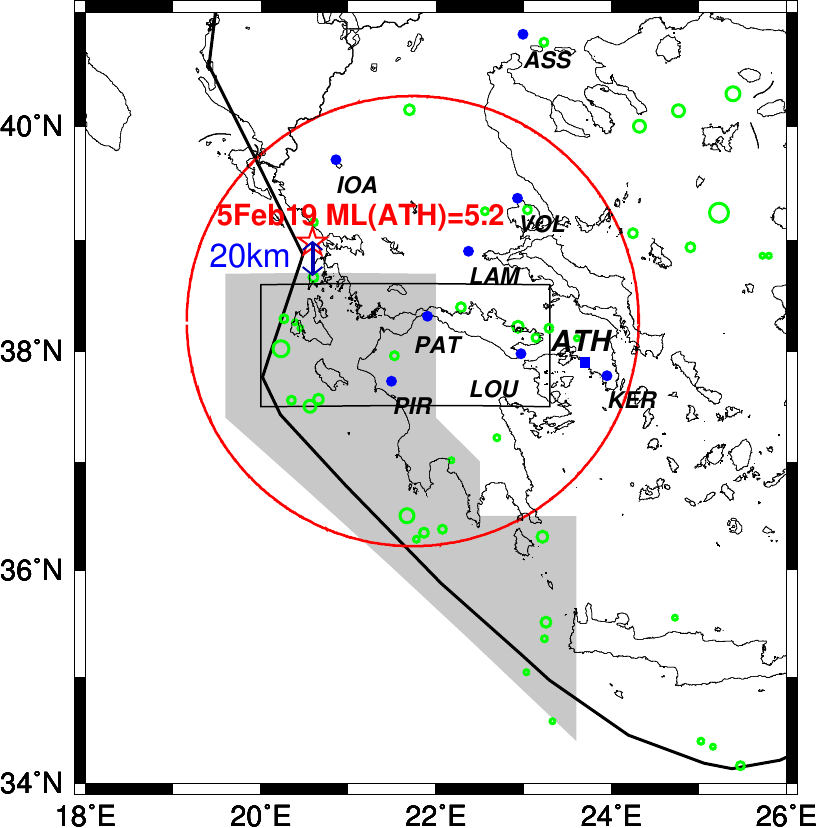}
\caption{Map depicting the epicenter (red star) of the ML(ATH)=5.2 EQ on 5 February 2019 located very close 
to the NorthWestern edge of the PIR selectivity map. \label{addf1}}
\end{figure}

\begin{figure}
\centering
\includegraphics[scale=0.4]{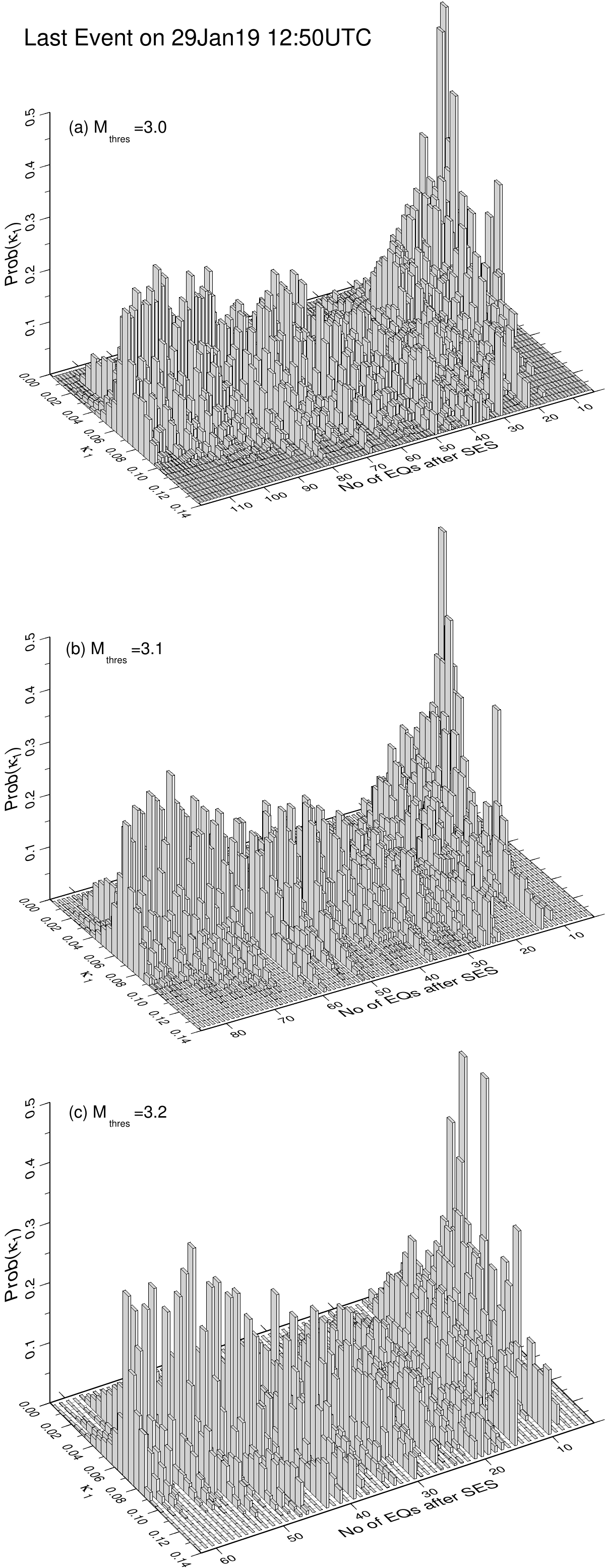}
\caption{(color online) (a) to (c): The probability distribution Prob$(\kappa_1)$ of $\kappa_1$ versus $\kappa_1$  as it results after the occurrence of each small EQ within the selectivity map of PIR (see the gray shaded area in  Fig. \ref{fig:1}) for 
various magnitude thresholds M$_{{\rm thres}}$=3.0, 3.1, and 3.2 after the SES activity at PIR on 9 January 2019. The last event considered is the ML(ATH)=3.5 EQ at 12:50 UTC on 29 January 2019 at 37.13$^o$N 20.59$^o$E. \label{addf1a} }
\end{figure}

\begin{figure}
\centering
\includegraphics[scale=0.6]{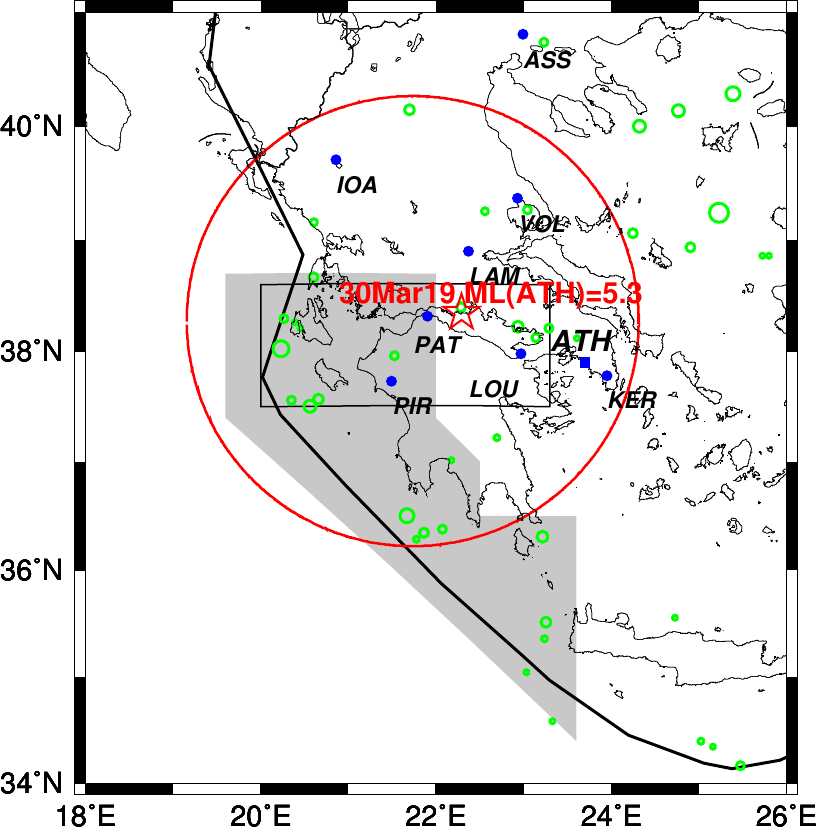}
\caption{Map depicting the epicenter (red star) of the ML(ATH)=5.3 EQ on 30 March 2019 located inside the PAT selectivity map (depicted by black rectangle which reproduces the rectangle with solid lines in 
Fig.8 of Ref.\cite{SAR08} mentioned in Section IV) at a distance around 
30km from the measuring station (PAT).\label{addf2}}
\end{figure}

\begin{figure}
\centering
\includegraphics[scale=0.4]{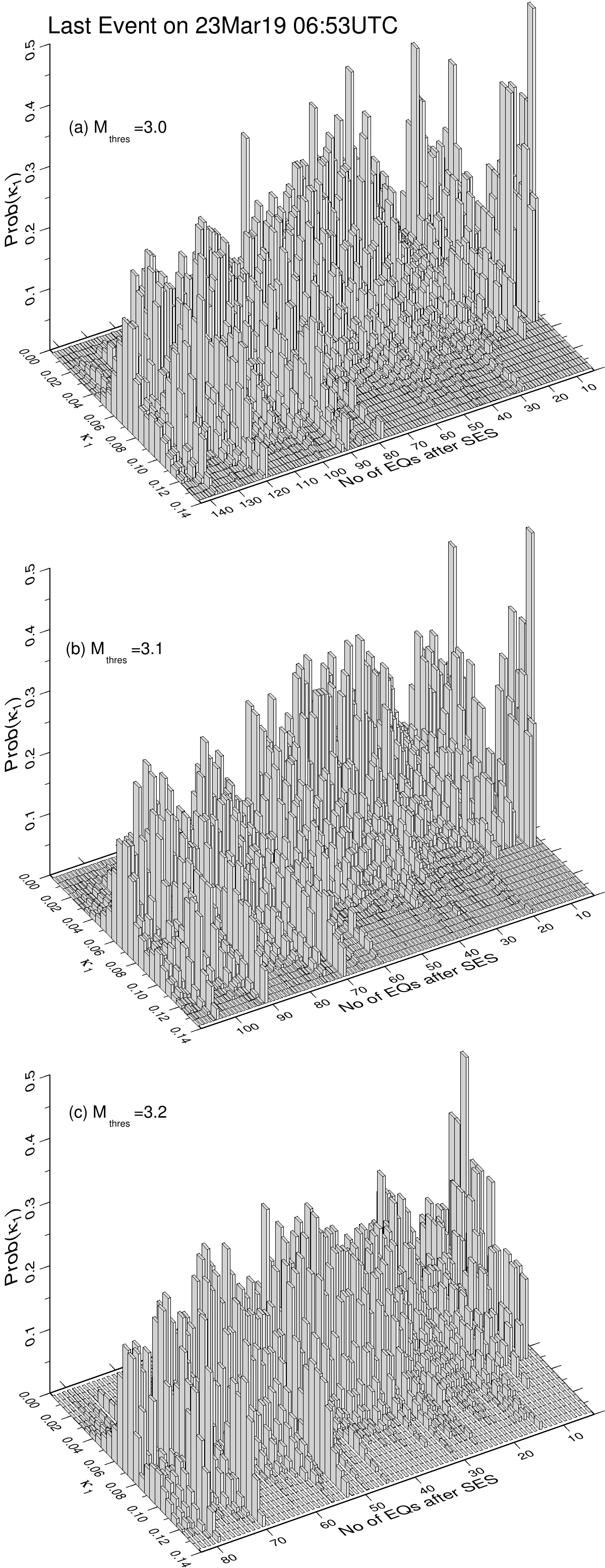}
\caption{(color online) (a) to (c): The probability distribution Prob$(\kappa_1)$ of $\kappa_1$ versus $\kappa_1$  as it results after the occurrence of each small EQ within the selectivity map of PAT 
(see the black rectangle in  Fig. \ref{addf2}) for 
various magnitude thresholds M$_{{\rm thres}}$=3.0, 3.1, and 3.2 after the SES activity at PAT on 3 January 2019. The last event considered is the ML(ATH)=3.2 EQ at 6:53 UTC on 23 March 2019 at 37.69$^o$N 20.61$^o$E. \label{addf2a} }
\end{figure}

\begin{figure}
\centering
\includegraphics[scale=0.4]{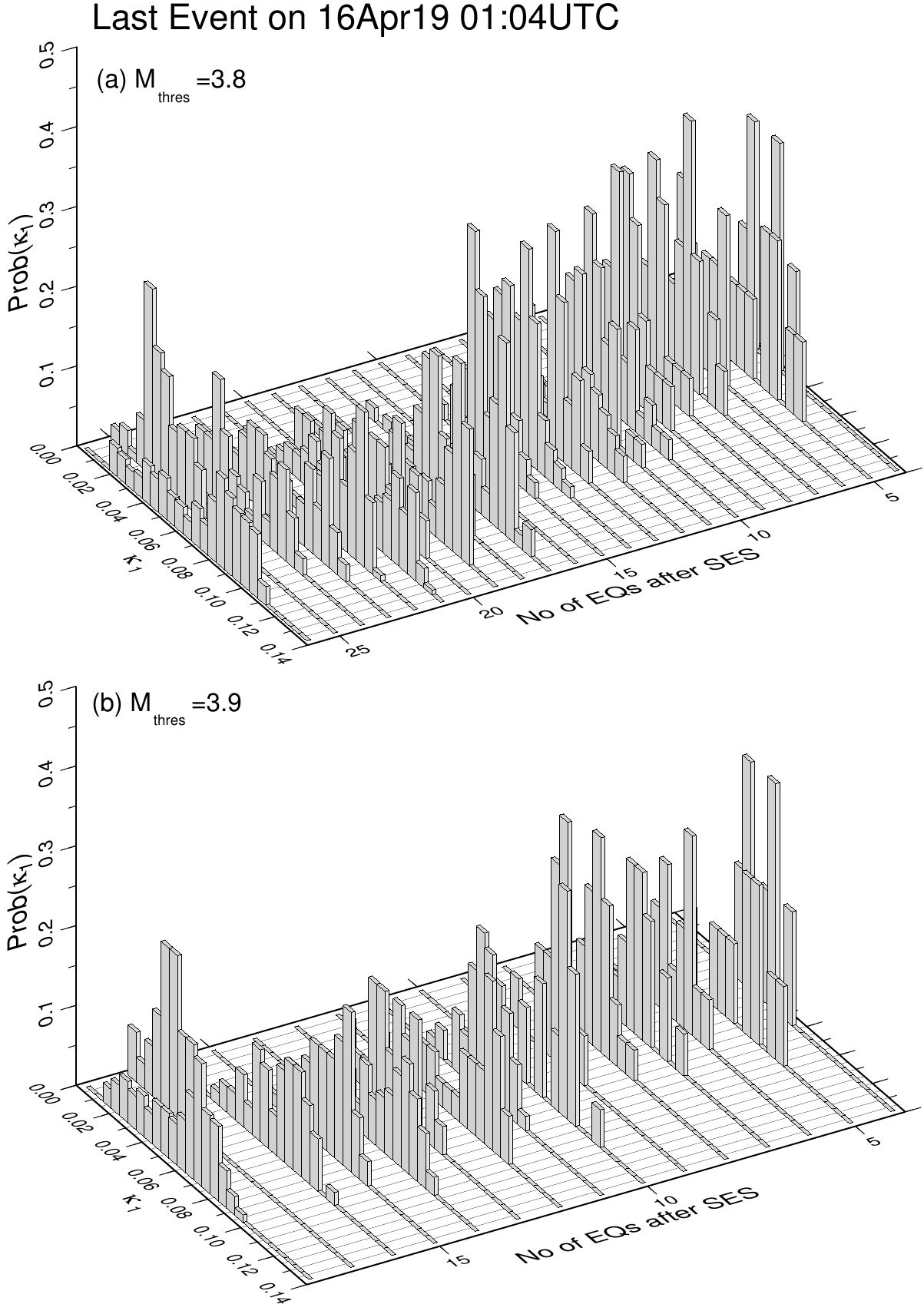}
\caption{(color online) The probability distribution Prob$(\kappa_1)$ of $\kappa_1$ versus $\kappa_1$  as it results after the occurrence of each small EQ within the selectivity map of PAT 
(see the black rectangle in  Fig. \ref{addf2}) for the  magnitude thresholds (a)M$_{{\rm thres}}$=3.8 and (b)M$_{{\rm thres}}$=3.9 after the SES activity at PAT on 3 January 2019. The last event considered is the ML(ATH)=3.9 EQ at 1:04 UTC on 16 April 2019 at 37.71$^o$N 20.71$^o$E. \label{addf2b} }
\end{figure}

\begin{figure}
\centering
\includegraphics[scale=0.6]{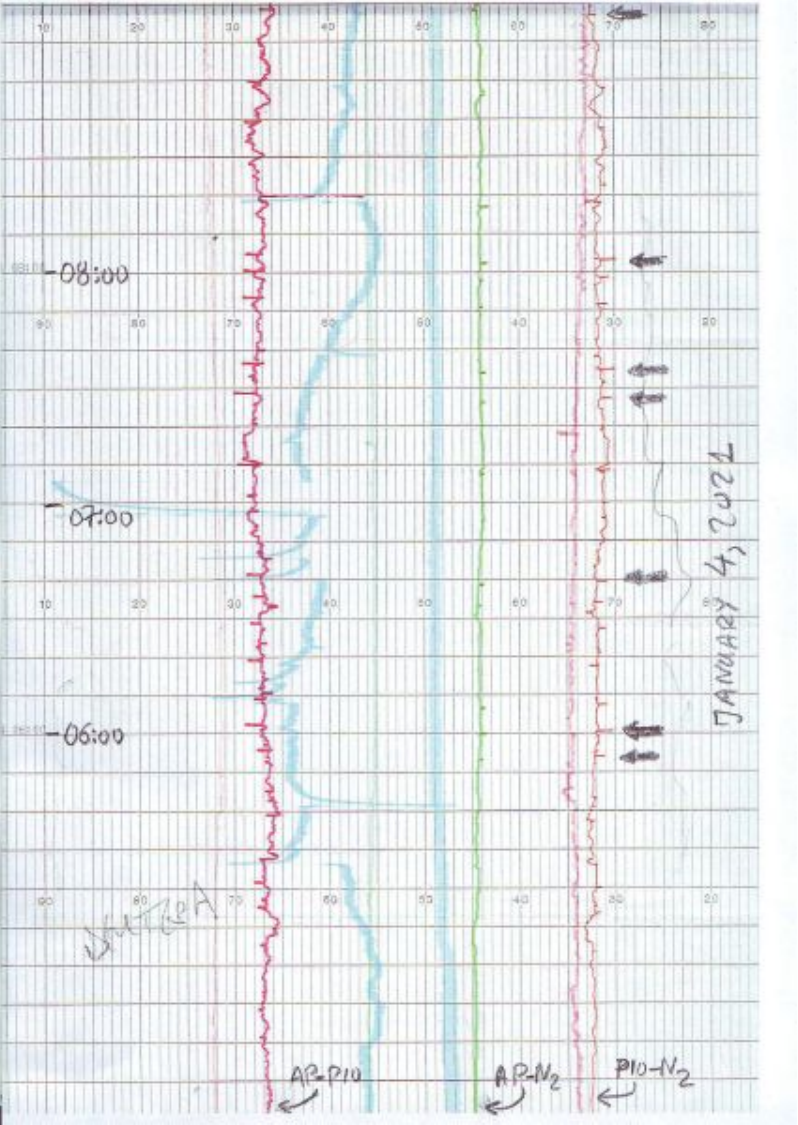}
\caption{{(color online) An excerpt of the raw data recordings at the central station of the telemetric network for the PAT
geoelectric station on 4 January 2021. The thick arrows show the most evident SES pulses at the 
three long measuring dipoles labeled AP-PIO, AP-N$_2$, and PIO-N$_2$.}   \label{fig:11} }
\end{figure}

\begin{figure*}
\centering
\includegraphics[scale=0.6]{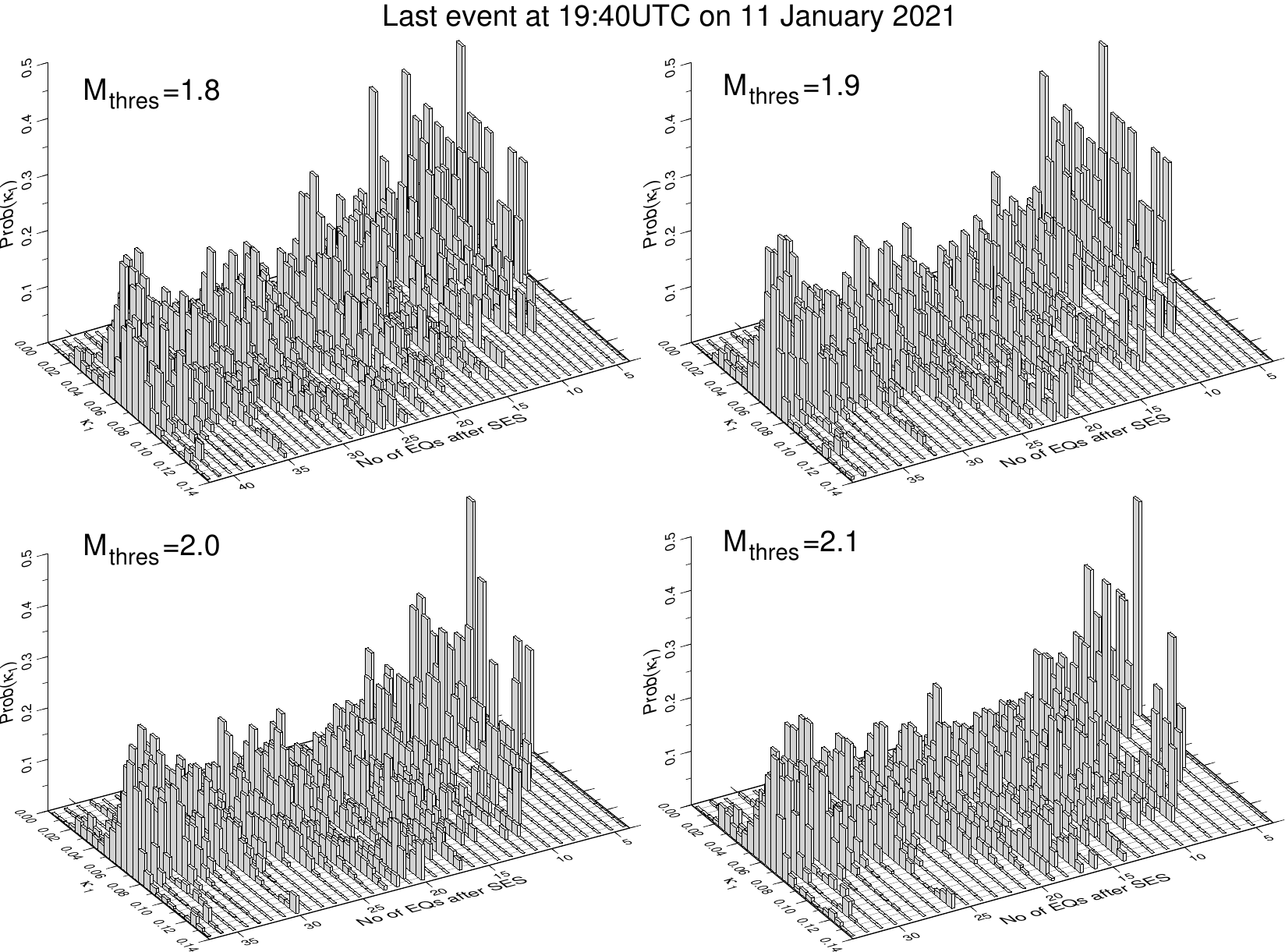}
\caption{{The probability distribution Prob$(\kappa_1)$ of $\kappa_1$ versus $\kappa_1$  as it results after the occurrence of each small EQ within the selectivity map of PAT 
(see the black rectangle in  Fig. \ref{addf2}) for various  low magnitude thresholds after the SES activity at PAT on 4 January 2021. The last event considered is the 
ML(ATH)=2.6 EQ at 19:40 UTC on 11 January  2021 at 38.39$^o$N 22.01$^o$E.}   \label{fig:12} }
\end{figure*}

\begin{figure}
\centering
\includegraphics[scale=0.22]{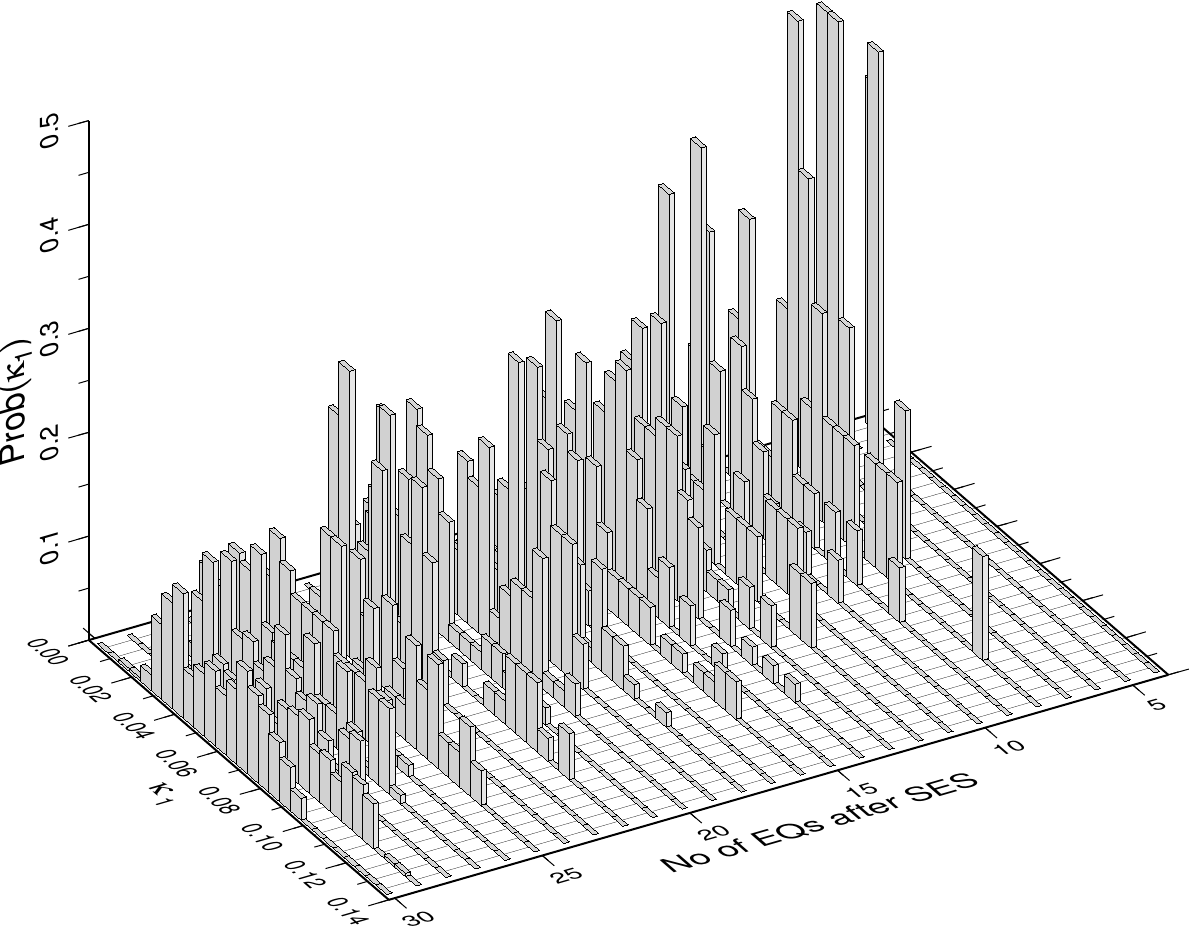}
\caption{{The probability distribution 
Prob$(\kappa_1)$ of $\kappa_1$ versus $\kappa_1$  as it results after the occurrence
of each small EQ within the selectivity map of PAT 
(see the black rectangle in  Fig. \ref{addf2}) for magnitude threshold 
M$_{{\rm thres}}=3.2$ after the SES activity at PAT on 4 January 2021. The last event considered is the 
ML(ATH)=3.2 EQ at 02:37 UTC on 17 February  2021 at 38.35$^o$N 21.94$^o$E.}   \label{fig:13} }
\end{figure}

\begin{figure}
\centering
\includegraphics[scale=0.32]{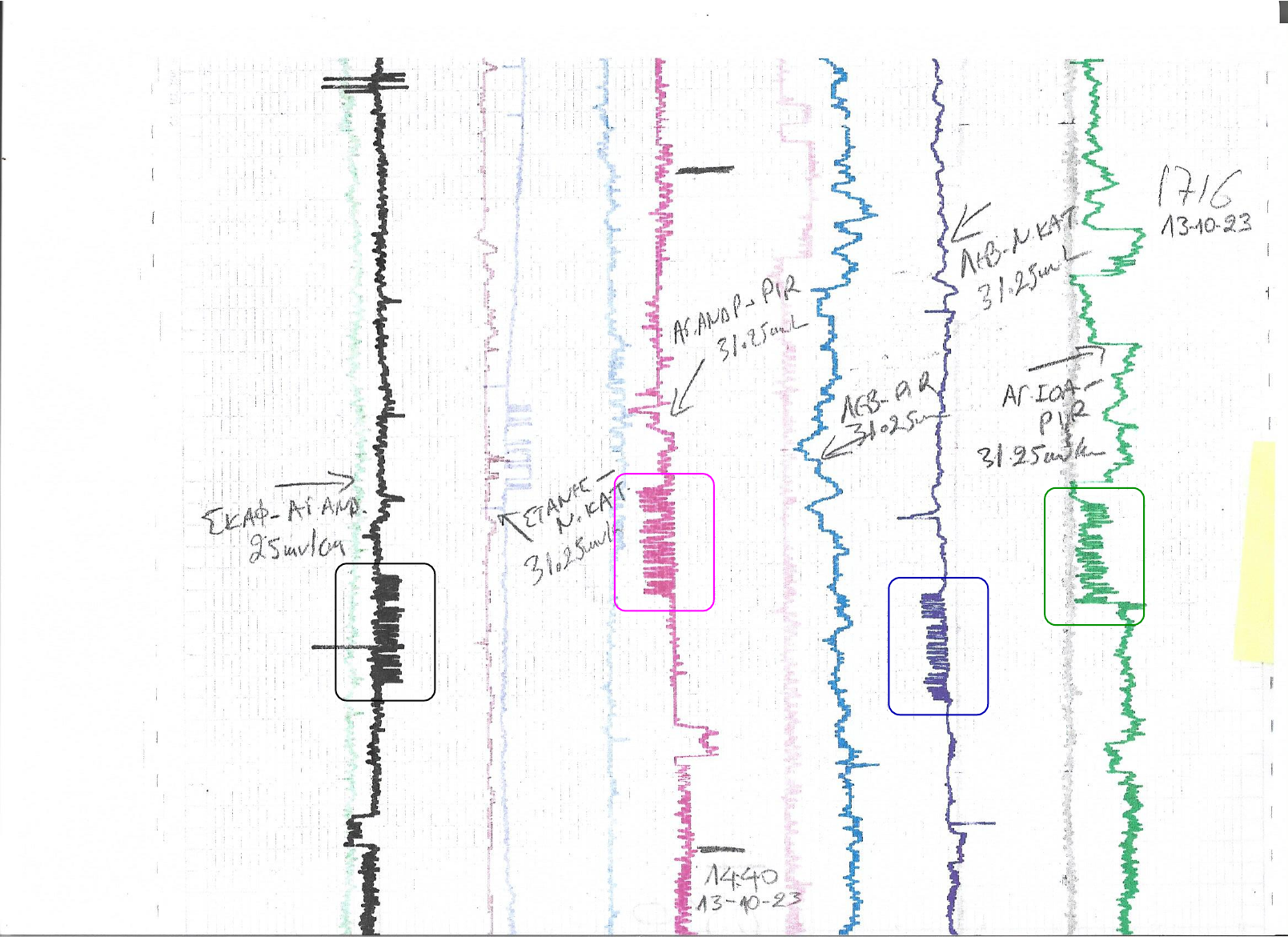}
\caption{An excerpt of the raw data recordings at the central station of the telementric network for the Pirgos (PIR) station on 13 October 2023. The simultaneous raw data recordings at four channels are surrounded by the relevant colored rectangles.}   \label{fig14} 
\end{figure}

\begin{figure}
\centering
\includegraphics[scale=0.32]{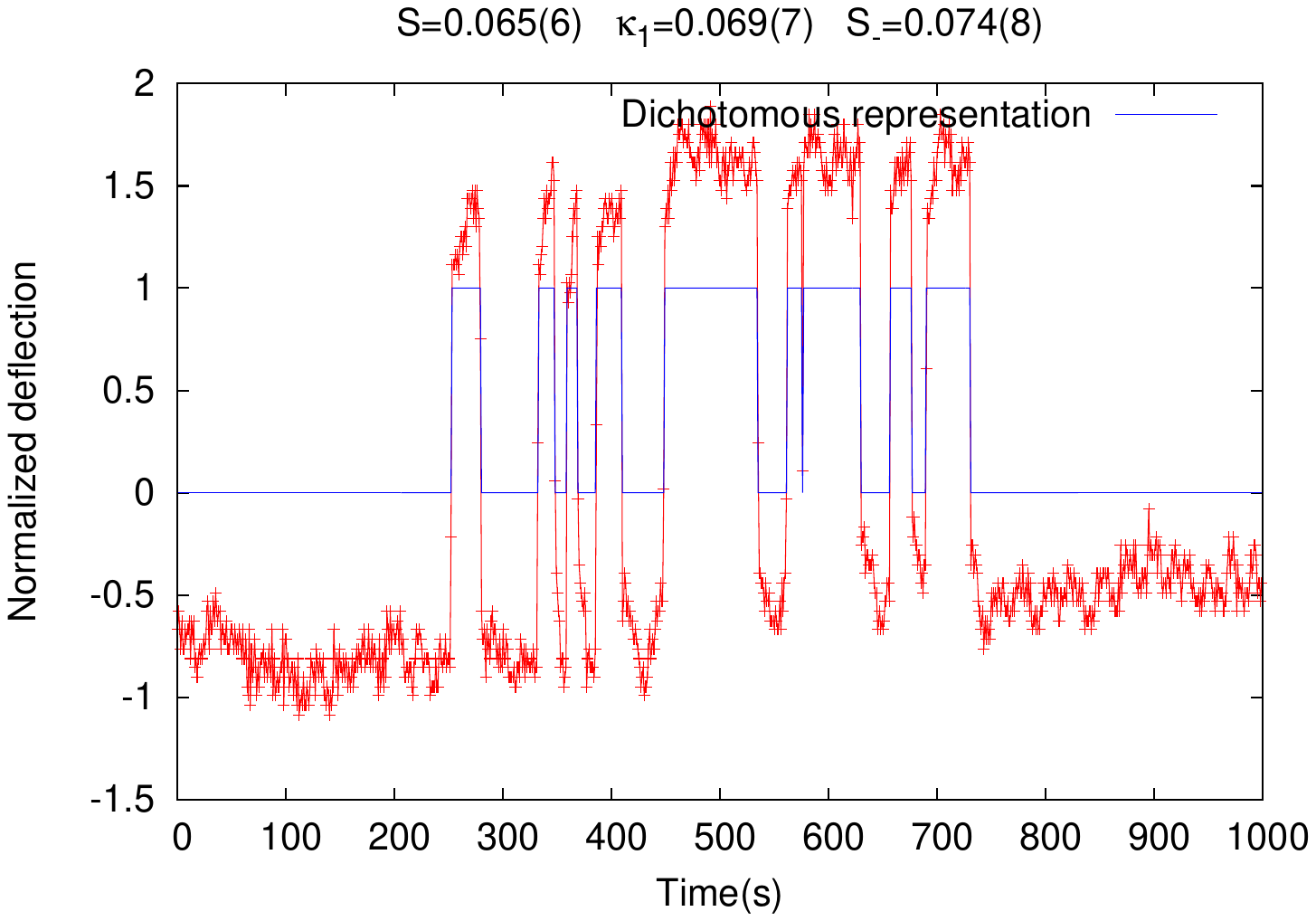}
\caption{The SES activity recorded at PIR on 6 July 2024.}   \label{fig15} 
\end{figure}

\vspace{6pt} 

\bibliographystyle{apsrev}

\end{document}